1# A Shift-insensitive Full Reference Image Quality Assessment Model Based on Quadratic Sum of Gradient Magnitude and LOG signals

Congmin Chen and Xuanqin Mou*Abstract*—Image quality assessment that aims at estimating the subject quality of images, builds models to evaluate the perceptual quality of the image in different applications. Based on the fact that the human visual system (HVS) is highly sensitive to structural information, the edge information extraction is widely applied in different IQA metrics. According to previous studies, the image gradient magnitude (GM) and the Laplacian of Gaussian (LOG) operator are two efficient structural features in IQA tasks. However, most of the IQA metrics achieve good performance only when the distorted image is totally registered with the reference image, but fail to perform on images with small translations. In this paper, we propose an FR-IQA model with the quadratic sum of the GM and the LOG signals, which obtains good performance in image quality estimation considering shift-insensitive property for not well-registered reference and distortion image pairs. Experimental results show that the proposed model works robustly on three large scale subjective IQA databases which contain a variety of distortion types and levels, and stays in the state-of-the-art FR-IQA models no matter for single distortion type or across whole database. Furthermore, we validated that the proposed metric performs better with shift-insensitive property compared with the CW-SSIM metric that is considered to be shift-insensitive IQA so far. Meanwhile, the proposed model is much simple than the CW-SSIM, which is efficient for applications.

*Index Terms*—Image quality assessment (IQA), full reference (FR), gradient magnitude, Laplacian of Gaussian, shift-insensitive.## I. INTRODUCTION

Since the distortion of information existed during the operation of image transmission, compression, restoration, etc., it is a significant procedure to evaluate the quality of digital images. In most cases, human beings are the conclusive observers who provide the ranks of image quality. Although the subjective score from the observation of human beings is able to estimate the image quality, automatic algorithms are absolutely much more convenient and economic in most practical applications. Image Quality Assessment (IQA) that computes the objective score in accordance with the subjective grading of distorted images has been widely applied to imitate the observation results of the human visual system (HVS).

Among different metrics, IQA models can be classified into three types: full reference (FR) metrics that needs reference image in evaluation, no reference (NR) or blind IQA that is effective where the pristine reference image is not available, and reduced reference (RR) IQA method that is employed when partial information of the source image is provided. This paper focuses on FR-IQA methods in which the reference image is assumed to be available and the quality to be perfect.

The traditional metrics such as the peak signal-to-noise ratio (PSNR) and the mean squared error (MSE), which computes image error pixel by pixel independently, are not well consistent with the subjective score of human judgement. The structural similarity (SSIM) index [1] was designed on the assumption that the HVS is sensitive to local structures and is able to capture the structural information when evaluating the quality of the visual signal. The multi-scale SSIM (MS-SSIM) [2], with an extension to single-scale algorithm, and the information weighted SSIM (IW-SSIM) [3], with consideration of the different types of local regions, bring better results than the original algorithm. The information fidelity criteria (IFC) [4] predicts the image quality by computing the information shared between the distorted images and reference images, and has been extended to a more efficient measurement named visual information fidelity (VIF) [5]. Another IQA metric based on the fact that HVS understands an image according to its low-level features is the feature-similarity (FSIM) [6] index, which uses the phase congruency (PC) to measure the significance of local structure and treats it as the primary feature. In consideration of the sensitivity of the image gradients to image distortions, the gradient magnitude (GM) similarity deviation (GMSD) [7] metric makes use of the variation of gradient based local quality map and shows excellent performance in image quality prediction.

A natural image contains a plenty of structures with various directional features and textures of higher level. However, any form of structures is consisted of a number of edges that is the primary structure in constituting any image. Hence high efficiently describing the edge structure is helpful to IQA model design. In fact, GM and LOG (Laplacian of Gaussian) are two fundamental operators to detect image edges and have played important roles in various IQA model designs [6]-[21]. GM removes the first-order statistics of the image (average of the luminance) and acquires intensity variation information, while the LOG removes the first- and second-order (contrast of the

Congmin Chen and Xuanqin Mou are with the Institute of Image Processing and Pattern Recognition, Xi'an Jiaotong University, Xi'an 710049, China (e-mail: chencongmin@stu.xjtu.edu.cn; xqmou@mail.xjtu.eud.cn).

luminance) information of the image [22] and retains semantic structures that consist of multi-layers structural information, i.e., from the low level directional features to more and more complex multi-layer textural features. When the GM and LOG are used to depict an edge structure, the GM contains mostly the contrast information while the LOG represents the phase information that demonstrates the sloping strength of the intensity variance, of which the zero-crossing (ZC) detection locates the image edge point. The joint distribution of the GM and LOG for natural images shows concordant relationship and can be used to produce an excellent blind IQA index [17]. However, the association of the GM and LOG has not been investigated in design of FR-IQA model, which leaves an interesting problem that whether or not we can exploit the relationship between GM and LOG and develop a FR-IQA with one or more possibly extraordinary property, even though either of GM and LOG has already been validated to be an efficient feature in FR-IQA model design.

Most existing FR-IQA models work well in the premise of that the reference and distortion images are well registered. This fact does not impede effectiveness of existing FR-IQA models because that in most FR-IQA applications the reference and distortion images is aligned. However, at present FR-IQA has been applied for various scenarios not only limited in evaluation of distortion images. Indeed, FR-IQA metrics can be used as the objective function of image denoising, compression, reconstruction, etc., in which the images to be evaluated are not always well-registered [23]. For example, in the evaluation algorithm of cameras for mobile communication devices and monitoring facilities, the moving scenes and the changing environments may cause mismatch between the reference and distortion images. To this regard, the registration precondition is indeed very strict for applications of FR-IQA models.

In recent years, deep neural networks (DNN), especially deep convolutional neural networks (CNN) have been being rapidly developed in various fields, mostly in the area of image processing [24]-[29]. In order to optimize parameters of the network, various loss functions are proposed for different networks to quantify and minimize the difference between the predicted result and the expected value in training procedure to adjust the parameters [30]. Mean square error (MSE) is the most conventionally used loss function. However, FR-IQA models, such as SSIM that captures perceptual features in the image while regardless of perceptually meaningless disturbances has been used to build SSIM-based loss functions. Such IQA model based functions has been validated to lead better results by improving presentation of perceptual structures [31]-[33].

In fact, either of MSE or perceptual aware metric is computed generally based overall pooling of point-wise error measures in the image, which means that the loss function is indeed computed based on a registered pattern. However, CNNs' success partly derives from the shift-invariance property of the CNN architecture which is mainly based on two strategies, i.e., shift-invariant convolution with sliding-window over the image, as well as down-sampling and pooling strategy, especially the max pooling operation [34]-[36]. In fact, existing CNNs cannot well handle the shift-invariance problem due to aliased representations when displacement is occurred [37], which might impede the generalization ability of CNNs. Along with this regard to loss function design, a shift-insensitive error measure between two image features, especially the primary edge features, would help to make better shift-invariance property of the CNN that might have improved generalization ability because shift-insensitive measure focuses on image contents while ignores feature positions. Randomness in feature positions would increase uncertainty of data while a shift-insensitive measure would weaken side effects caused by the uncertainty.

At present, there is hardly any work on shift-insensitive FR-IQA model development except the complex wavelet SSIM (CW-SSIM) metric [38]. By making use of the phase information, the key idea of CW-SSIM index is that small geometric distortions in images lead to consistent phase changes in the local wavelet coefficients, and a consistent phase (position) shift of the coefficients does not change the structural content. The CW-SSIM metric has been proved to be sensitive to structural distortions and insensitive to position shift. Such method works robustly on images with small translations without a pre-processing step for image registration. However, CW-SSIM extracts copious Gabor features that span over different directions, angles, scales, and phase parameters, and has a complex computation model, which may hinder its practical applications, including possible action as the loss functions of CNNs. In the same time, the Gabor filters are featured with angle favour, which would prefer to artifact structures in specific directions. By contrast, both the GM and LOG is circular symmetric operator without any angle favour. If we can design a shift-invariant FR-IQA model based on the two operators, this symmetric property would benefit the designed model being simple and unblemished.

In this study, we find that based on the concordant relationship in depicting image edges between the GM and LOG, the quadratic sum of the GM and LOG demonstrates a shift-insensitive property in representing image edge features. And the arithmetical difference between the quadratic sums of the reference and distortion images shows to be a very simple and efficient FR-IQA model that is superior to most state-of-the-art models in terms of accordant predication with human opinion. At the same time, the proposed model has better shift-invariant performance compared to CW-SSIM in terms of predicting perceptual quality of distorted images.

The remainder of this paper is organized as follows. Section II shows the definitions in this paper and the arithmetic of the modelling process in detail. Section III reports our experimental results, comparison and analysis. Finally, we conclude the paper in Section IV.

## II. METHODS

Since the GM is sensitive to the change of the image intensity, and the LOG signal is sensitive to the location of edges, we use a combination of LOG and GM signals of the image as the IQA feature to express the local texture. In this section, we first introduce the GM and LOG functions. Then, we propose to use the root of the sum square of GM and LOG signals as the quality feature, and test the feature on the

Heaviside step function, which can be used to represent an ideal edge in natural images along with any direction. We find that with this simple combination of the GM and LOG signals, the model is able to work more stably and could ignores the location change of the edge. The last paragraph shows an analysis on that the function stays invariable near the edge when an appropriate parameter is provided, thus we suggest that the proposed feature can work stably when small spatial translation occurs in the reference or distortion images.

*A. LOG Signal and Gradient Magnitude*

Laplacian filter is a second-order derivative filter used in edge detection. Since the derivative filters are sensitive to noise, a Gaussian filter is commonly applied to smooth the noise. This two-step process is called the Laplacian of Gaussian, or LOG. The LOG operator takes the second derivative of the given signals. When the image is basically uniform, the LOG will give zero. Wherever a change occurs, the LOG will give a positive response to the darker side and a negative response to the lighter side, where the zero-crossing point represents position of the edge.

The LOG filter is defined as:

$$\boldsymbol{h}_{LOG}(x,y|\sigma) = -\frac{1}{\pi\sigma^4}(1-\frac{x^2+y^2}{2\sigma^2})e^{-\frac{x^2+y^2}{2\sigma^2}} \quad (1)$$

where the variables $x$ and $y$ denote the coordinate of the input image, parameter $\sigma$ represents spatial scale factor of the LOG filter. We denote an image by $\boldsymbol{I}_R$ and a distorted image by $\boldsymbol{I}_D$, then the LOG map can be computed as:

$$L_{R,\sigma} = \boldsymbol{I}_R \otimes \boldsymbol{h}_{LOG}(x,y|\sigma) \quad (2)$$

$$L_{D,\sigma} = \boldsymbol{I}_D \otimes \boldsymbol{h}_{LOG}(x,y|\sigma) \quad (3)$$

We use $L_{R,\sigma}$ to denote the transformation results with the parameter $\sigma$ of the reference images, while $L_{D,\sigma}$ to denote the filtered distorted signals.

The image gradient magnitude that is defined as the root mean square of image directional gradients along two orthogonal directions, such as $x$ and y axes, has been applied to extract the edge information in different ways [6]-[11]. In order to suppress image noise, a Gaussian filter is also used before the convolving process. Hence, the first-order derivative of Gaussian filters on horizontal direction and vertical direction are defined as:

$$\boldsymbol{h}_x(x,y|\sigma) = (-\frac{1}{2\pi\sigma^4})xe^{-\frac{x^2+y^2}{2\sigma^2}} \quad (4)$$

$$\boldsymbol{h}_y(x,y|\sigma) = (-\frac{1}{2\pi\sigma^4})xe^{-\frac{x^2+y^2}{2\sigma^2}} \quad (5)$$

We convolve the reference image with the two directional derivative filters to yield the horizontal and vertical gradient images:

$$\boldsymbol{d}_{R,x,\sigma} = \boldsymbol{I}_R \otimes \boldsymbol{h}_x(x,y|\sigma) \quad (6)$$

$$\boldsymbol{d}_{R,y,\sigma} = \boldsymbol{I}_R \otimes \boldsymbol{h}_y(x,y|\sigma) \quad (7)$$

The GM of reference images is computed as follows:

$$D_{R,\sigma} = \sqrt{\boldsymbol{d}_{R,x,\sigma}^2 + \boldsymbol{d}_{R,y,\sigma}^2} \quad (8)$$

where σ denotes the parameter of scale in the Gaussian filter. The GM of distorted images can be produced in the same way, as in (9):

$$D_{D,\sigma} = \sqrt{\boldsymbol{d}_{D,x,\sigma}^2 + \boldsymbol{d}_{D,y,\sigma}^2} \quad (9)$$

To further remove contrast variation in the image in large scale, we apply divisive normalization process [16] to the LOG and GM signals. The divisive normalization processes are shown as following equations:

$$U_{R,\sigma}(i,j) = \frac{k \cdot L_{R,\sigma}(i,j)}{\sqrt{G_{2\sigma}(i,j) * [D_{R,\sigma}^2(i,j) + k^2 L_{R,\sigma}^2(i,j)] + c_0}} \quad (10)$$

$$V_{R,\sigma}(i,j) = \frac{D_{R,\sigma}(i,j)}{\sqrt{G_{2\sigma}(i,j) * [D_{R,\sigma}^2(i,j) + k^2 L_{R,\sigma}^2(i,j)] + c_0}} \quad (11)$$

where $k$ is an ratio factor to adapt magnitude difference between the LOG and GM signals, $G_{2\sigma}$ denotes the normalization Gaussian filter with the spatial scale factor that is 2 times of the GM and LOG scale, and $c_0$ is a constant that makes the denominator not to be zero. Furthermore, $U_{D,\sigma}$ and $V_{D,\sigma}$ can be produced from distorted images in the same way.

Fig. 1 shows the relationship between GM and LOG through the filtered results of a 1-D Heaviside step function $u_0(x)$ in Fig. 1(a). We use Gaussian filter $G(x)$ to smooth the step signal first, as shown in Fig. 1(b), and calculate the first-order and second-order derivatives of the edge, denoted by $d_1(x)$ and $d_2(x)$, as shown in Figs. 1(c) and 1(d), respectively. Then the joint distribution of $d_1(x)$ and $d_2(x)$ is given in Fig. 1(e). Three arrows in different colors indicate the corresponding points of edge phase respectively. According to the variation of phase and amplitude of $d_1(x)$ and $d_2(x)$, the two curves hold a harmonious relation near the edge point, as shown in Fig. 1(e). The last curve $R(x)$ is the root of the quadratic sum of $d_1(x)$ and $d_2(x)$ which shows a property of responding to an edge function but insensitive to the exact position of the edge, as shown in Fig. 1(f). To be specific, the insensitivity can be observed by comparing the Figs. 1(f) and 1(b), where in Fig. 1(b), the scale-filtered edge signal has a sharpen profile compared to the former.

Since GM and LOG are relatively complemented in image edge representation and can be used in BIQA model design, in this study we design a simple combination of $U_{R,\sigma}$ and $V_{R,\sigma}$ to express the shift-insensitive structural information of reference images, defined as the quadratic sum of the normalized GM and LOG signal (QGL). The QGL feature for the reference image is calculated as:

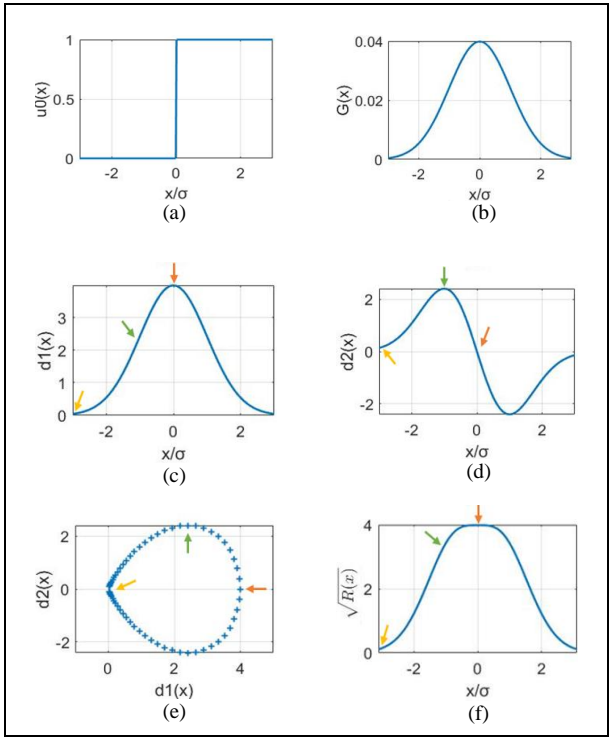

Fig. 1. Step function and its filtering results. The colorful arrows point out the corresponding locations on different curves. (a) Step function. (b) Gaussian function applied to smooth the edge. (c) First-order derivative of the edge. (d) Second-order derivative of the edge. (e) Scatter plots of *d1(x)* versus *d2*(x). (f) Root of the quadratic sum of *d1(x)* and *d2(x)*.

$$\boldsymbol{q}_{R,\sigma}(i,j) = \sqrt{U_{R,\sigma}^2(i,j) + V_{R,\sigma}^2(i,j)} \quad (12)$$

In the same way, the QGL feature for the distortion image can be computed as:

$$\boldsymbol{q}_{D,\sigma}(i,j) = \sqrt{U_{D,\sigma}^2(i,j) + V_{D,\sigma}^2(i,j)} \quad (13)$$

which is generated from distorted images.

Fig. 2 shows the gray-scale maps of the LOG, GM, and the $r_{R,\sigma}$ generated from three reference images in LIVE database. By observation, the contours of the $r_{R,\sigma}$ image show more distinct and bold edges that hints shift-insensitive property of edge representation.

### B. The Proposed FR-IQA Model

We use the similarity computation to ascertain the structural difference between distorted and reference images. The similarity map is defined as:

$$Q(i,j) = \frac{2\boldsymbol{q}_{R,\sigma}(i,j)\boldsymbol{q}_{D,\sigma}(i,j) + c_1}{\boldsymbol{q}_{R,\sigma}^2(i,j) + \boldsymbol{q}_{D,\sigma}^2(i,j) + c_1} \quad (14)$$

where $c_1$ is a positive constant that supplies numerical stability. Eq. (14) measures the local similarity at each image pixel. To yield the overall evaluation score of the image, a polling strategy shall be introduced to integrate the similarities of all image pixels.

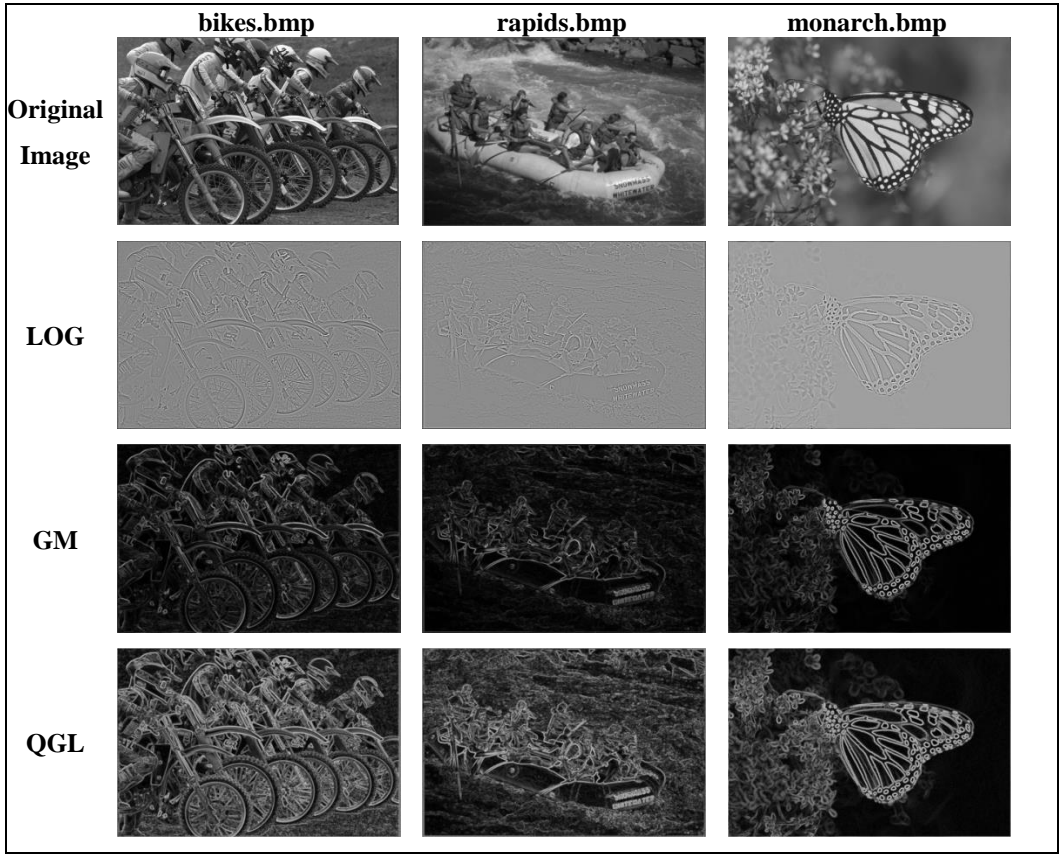

Fig. 2. The gray-scale maps of the LOG signal, gradient magnitude, and the $r_{R,\sigma}$ generated from three reference images in LIVE database.

Although researchers have made many efforts to design weighted pooling methods [3], [43]-[47], the computation of weights is not always significant [48]. Average pooling method supposes that each part of the image is of equal importance to the overall quality. It is a general way to produce the general estimate of the image quality [44], [45]. Based on this, in this study we propose the first FR-IQA metric named as the mean of the QGL signal, abbreviated as mQGL, as shown in Eq. (15):

$$mQGL = \frac{1}{N}\sum_{i,j} Q(i,j) \qquad (15)$$

In the same time, standard deviation pooling has been validated to be a more efficient method in synthesizing for gradient similarity based IQA method [7], here we accordingly propose the second FR-IQA metric named as the standard deviation of the QGL signal, abbreviated as sQGL, as shown in Eq. (16):

$$sQGL = \sqrt{\frac{1}{N}\sum_{i,j}(Q(i,j) - mean(Q))^2} \qquad (16)$$

Note that the value of mQGL means the average similarity between reference and distorted images, which gives higher score to higher image quality, while sQGL indicates the difference between reference and distorted images, and gives higher score to the larger distortion level and lower image quality.

*C. Analysis on the Shift-insensitive Property*

In this section we analyze the shift-insensitive property of the proposed QGL feature and explore the parameter selection problem in the QGL feature by using 1-D Heaviside step signal. Consider a zero-mean Gaussian function $G(x)$ with scale factor $\sigma$, which is defined as:

$$G(x) = \frac{1}{\sqrt{2\pi}\sigma}e^{-\frac{x^2}{2\sigma^2}} \qquad (17)$$

We get the first and second derivative of the Gaussian function to imitate the gradient and LOG function on 1-D signals:

$$G_1(x) = -\frac{x}{\sqrt{2\pi}\sigma^3}e^{-\frac{x^2}{2\sigma^2}} \qquad (18)$$

$$G_2(x) = -\frac{1}{\sqrt{2\pi}\sigma^3}(1-\frac{x^2}{\sigma^2})e^{-\frac{x^2}{2\sigma^2}} \qquad (19)$$

The Heaviside step function $u_0(x)$ can be used to represent an ideal edge in natural images without considering its direction property, and the smoothing process by Gaussian filter is expressed as:

$$u(x) = G(x) * u_0(x) \qquad (20)$$

Thus the first and second derivatives of $u(x)$ are computed as follows:

$$d_1(x) = \frac{d[u(x)]}{dx} = G_1(x) * u_0(x) \qquad (21)$$

$$= \int_{-\infty}^{x} G_1(t)dt$$
$$= \frac{1}{\sqrt{2\pi}\sigma}e^{-\frac{x^2}{2\sigma^2}} = G(x)$$

$$d_2(x) = \frac{d[d_1(x)]}{dx} = G_2(x) * u_0(x)$$
$$= \int_{-\infty}^{x} G_2(t)dt \qquad (22)$$
$$= -\frac{x}{\sqrt{2\pi}\sigma^3}e^{-\frac{x^2}{2\sigma^2}} = G_1(x)$$

We define the sum of square of $d_1$ and $d_2$ as:

$$R(x) = d_1^2(x) + [k \cdot d_2(x)]^2$$
$$= \frac{\sigma^4 + k^2x^2}{2\pi\sigma^6}e^{-\frac{x^2}{\sigma^2}} \qquad (23)$$

where $k$ is a ratio factor to compensate the magnitude difference between $d_1(x)$ and $d_2(x)$. In the following, we will theoretically analyse how to determine the factor $k$.

Aiming at insensitivity on spatial translation, we hope $R(x)$ to be stable near the central point of the edge. Therefore, we denote the derivative of $R(x)$ by $f(x)$, and search an optimal $k$ when the absolute value of $f(x)$ is minimum:

$$f(x) = \frac{dR(x)}{dx}$$
$$= \frac{2x(k^2\sigma^2 - \sigma^4 - k^2x^2)}{\pi\sigma^8}e^{-\frac{x^2}{\sigma^2}} \qquad (24)$$

As is shown in Eq. (23), the function $R(x)$ is what we want as a shift-insensitive filter, in which the central point $R(0)$ is a constant with a given $\sigma$, regardless of the value of $k$. Since $R(x)$ is an even function and $f(0) = 0$, $R(0)$ is an extremum or a stagnation point. Noted that $R(0)$ should be a maximum, not a minimum point, so that the second-order derivative of $R(x)$ should not be larger than zero according to the relationship between the extreme value and derivative of function, that is

$$f'(0) \leq 0 \qquad (25)$$

Therefore, an inequality

$$k \leq \sigma \qquad (26)$$

can be obtained as a necessary condition of the filter. Since $f(x)$ is an odd function, the value $f(0)$ cannot be an extreme point. If and only if $f'(0) = 0$, $f(0)$ is a stagnation point, and the value of $k$ turns to be equal to $\sigma$.

For a more accurate optimization process of $k$, we suggest that $R(x)$ varies slowly in $[-t, t]$, and a wide range of the flat region is expected, thus the derivative of $R(x)$ is also expected to be stable. In an ideal situation when $t$ is very small, let

$$f(t) = 0, and\ f'(t) = 0 \qquad (27)$$

Thus, two expressions of $k$ can be written as follows:

$$k^2 = \frac{\sigma^2}{1-\beta^2}, \qquad (28)$$

$$k^2 = \frac{\sigma^2(1-2\beta^2)}{2\beta^4 - 5\beta^2 + 1} \quad (29)$$

where $\beta = t/\sigma$ is employed as a ratio coefficient for different scales of Gaussian function. Fig. 3 shows the two curves demonstrating the ideal $k$ from $\beta = 0$ to $\beta = 1$. However, $f(t)$ and $f'(t)$ are impossible to be exactly equal to zero, such $k$ curves are ideal values for each individual point that are hoped to be approached.

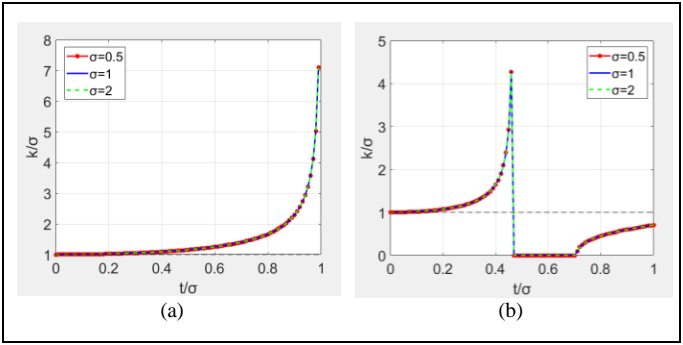

Fig. 3. The ideal k value computed from Eq. (28) and Eq. (29). The value has been divided by σ for normalization on different scales.

It should be noted that the valid width of the filtering window is usually within [-σ, σ], thus it is more important to find an appropriate $k$ for smaller $\beta$. As is mentioned in Eq. (26), $k \leq \sigma$ is a necessary condition, hence $k = \sigma$ is the only optimal value that is the closest to the ideal curve. Note that for 2-D signals, the value of $k$ should be multiplied by $\sqrt{2}$.

Fig. 4 shows the root of function $R(x)$ on different scales ($\sigma = 0.5, 1, 2$) with parameter $k = \sigma$ in the computation. The value of $R(x)$ has been normalized for any $\sigma$. This figure reveals that an optimal $k$ leads to a flat top in function $R(x)$, where the filtration result gives similar values and declines slowly within an interval near the edge position. Therefore, the proposed feature is capable to respond to an edge function but insensitive to the edge position within a spatial translations.

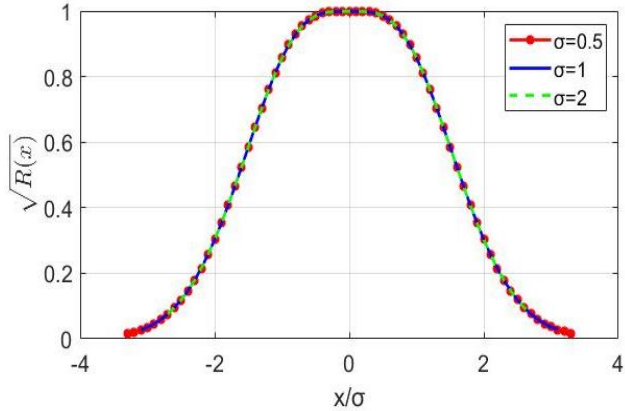

Fig. 4. Root of $R(x)$ on three different scales when $k = \sigma$ in the calculation process.

## III. EXPERIMENTAL RESULTS

### A. Test Database

We evaluate the performance of the proposed mQGL and sQGL on three publicly accessible IQA databases: LIVE [39], CSIQ [40], and TID2013 [41]. The LIVE database consists of 779 distorted images created from 29 reference images with 5 types of distortions: JPEG compression, JPEG2000 compression, white noise, Gaussian blur and simulated fast fading. The CSIQ database contains 866 subject-rated distorted images which are created from 30 reference images with 6 different types of distortions: JPEG compression, JPEG2000 compression, additive white noise, additive pink Gaussian noise, Gaussian blur, and global contrast decrements. The Difference Mean Opinion Score (DMOS) values are provided as part of the LIVE database and CSIQ database. For each image, higher DMOS value means higher distortion and lower image quality in the subjective evaluation. The TID2013 database is the largest database of the commonly used databases which is intended for evaluation of full-reference image visual quality assessment metrics. It contains 3000 distorted images, generated from 25 reference images with 24 types of distortions at 5 levels. This database covers the most types of distortions among all existed databases. The human subjective score has been given as Mean Opinion Score (MOS) in TID2013 database, where higher MOS value means higher subjective image quality. The distortion types in these databases reflect a broad range of image impairments, and are widely applied in IQA research.

The competitors for evaluating the proposed methods are selected from the state-of-the-art FR-IQA models. More specifically, SSIM is a successfully influential model that predicates image quality by measuring structural similarity. MS-SSIM and IW-SSIM are derived from SSIM. VIF and FSIM methods are both outstanding FR-IQA methods that are extensively accepted and used in the field of image processing. The manifestations of GMSD and NLOG methods are taken into account for the performance when only GM or LOG is applied in feature extraction. Meanwhile, conventionally used PSNR is also included as the baseline.

In the experiment, the scale factor $\sigma$ of the Gaussian filter is selected as 0.5, which shows the best performance in image quality estimating, and the scale factor for the divisive normalization Gaussian filter is $2\sigma$. The constant $c_0$ in the divisive normalization process is selected as 1, while the constant $c_1$ for the quality map computation is selected as 0.0009. The constants are selected experimentally for achieving the best IQA performance. However, the selection is not very sensitive to influence the model performance by our experimental observations. In evaluating shift-insensitive performance of the proposed method, we use with the CW-SSIM as the competitor because that it is the only one shift-insensitive FR-IQA metric so far. In the evaluation, we use the scale factor $\sigma = 1$ to address the shift-insensitive property of the proposed models.

### B. IQA Performance

One of the most commonly used performance metrics, SROCC (Spearman rank-order correlation coefficient) index that measures the monotonicity of the computed result and the subjective score, is employed to evaluate the performance of the proposed IQA model, as well as the competitors. We investigate the results of the model scores for each image from the three benchmark databases, which include both pristine images and distorted images, to compare with the value of DMOS or MOS value given from the databases. The SROCC regarding to the model scores versus the human subjective opinion scores on the different databases are shown in TABLE 1. The top three models for each database are shown in boldface.

As shown in TABLE 1, the proposed sQGL with deviation pooling strategy ranks 1$^{st}$ on TID2013, 2$^{nd}$ on CSIQ within an ace of the best one, and.4$^{th}$ on LIVE databases. Meanwhile, mQGL with average pooling strategy is comparatively weak. Standard deviation pooling shows more efficient property in the proposed QGL model. This experimental result indicates that the proposed metric has stable performance on different databases, and can be efficient on different types of distorted images. A model working well on TID2013 is much more significant since the TID2013 database is the largest database among the three commonly used databases. Meanwhile, the weighted average achieved by sQGL ranks 1$^{st}$ across the three databases, which shows that the proposed method works stably and robustly on a comprehensive range of images. According to this comparison, the proposed sQGL model achieves stable performance across the three databases, especially on the largest TID2013 database, which is significantly better than all the other competitors.

TABLE 1
PERFORMANCE COMPARISON OF DIFFERENT FR-IQA MODELS ON THREE BENCHMARK DATABASES.
THE TOP THREE MODELS FOR EACH DATABASE ARE SHOWN IN BOLDFACE.

| SROCC | LIVE (779 images) | CSIQ (866 images) | TID2013 (3000 images) | Weighted Average |
|---|---|---|---|---|
| PSNR | 0.8756 | 0.8058 | 0.6394 | 0.7100 |
| SSIM [1] | 0.9479 | 0.8756 | 0.7417 | 0.8012 |
| MS-SSIM [2] | 0.9513 | 0.9133 | 0.7859 | 0.8374 |
| IW-SSIM [3] | 0.9567 | 0.9213 | 0.7779 | 0.8346 |
| IFC [4] | 0.9259 | 0.7671 | 0.5390 | 0.6463 |
| VIF [5] | **0.9636** | 0.9195 | 0.6770 | 0.7703 |
| FSIM [5] | **0.9634** | 0.9240 | **0.8022** | **0.8519** |
| NLOG-MSE | 0.9405 | 0.9259 | 0.7734 | 0.8299 |
| NLOG-COR | 0.9429 | **0.9308** | 0.7772 | 0.8336 |
| GMSD [6] | **0.9603** | **0.9570** | **0.8044** | **0.8590** |
| RFSIM [42] | 0.9438 | 0.9292 | 0.7744 | 0.8317 |
| mQGL | 0.9524 | 0.9227 | 0.7903 | 0.8422 |
| sQGL | 0.9574 | **0.9550** | **0.8103** | **0.8619** |

In order to further test the performance on different distortion types, TABLE 2 shows the performance of the proposed model and the competitors on each individual distortion type, where the top three models for each distortion type are shown in boldface. The experimental result of the comparison reveals that the proposed sQGL performs well on most distortion types, which also reveals robust property of the proposed method on different image distortion types across the three databases.

### C. Test on the Shift-insensitive Property

We transform the reference images in LIVE database by spatial shift with the range from 0 to 10 pixels on horizontal and vertical directions, and investigate the SROCC between the image subjective scores and model scores computed by the proposed model in association with displacement distance. The following metrics, GMSD [7], SSIM [1] and CW-SSIM [38] are used as the competitors. The experimental results on LIVE database are shown in Fig. 5. The horizontal ordinate denotes the distance of spatial translation in the image, and the vertical ordinate denotes the SROCC between the objective scores computed by different FR-IQA metrics and the subjective scores provided in the database. In this experiment, as mentioned before, we select a larger scale factor in the Gaussian filter for the sake of detection on the shifted edge.

Experimental results show that the proposed model performs better than the competitors when small spatial translation occurs in the image, and the SROCC value of the average pooling method mQGL declines more slowly than sQGL with deviation pooling. We mark the value of mQGL on the end of the curve where the translation distance is 10 in the figure. Although the SROCC of mQGL is not the highest on the point with zero displacement, the value falls most slowly among these metrics. When the reference image is shifted by 5 pixels, the SROCC value of mQGL holds upon 0.7 along the two directions, while GMSD drops below 0.5, SSIM falls below 0.2, and CW-SSIM falls below 0.65, given that CW-SSIM was designed for shift-insensitivity purpose. At the end of the curve, only the proposed model holds the value of SROCC larger than 0.6.

The comparisons on CSIQ and TID2013 databases are shown in Fig. 6 and Fig. 7 separately. Those comparisons also indicate that the proposed models ignore the spatial position of an edge in a local area, thus tends to be shift-insensitive when small spatial shift occurs in images.

TABLE 2
PERFORMANCE OF COMPARISON OF THE IQA MODELS ON EACH INDIVIDUAL DISTORTION TYPE IN TERMS OF SROCC.
THE TOP THREE MODELS FOR EACH TYPE ARE SHOWN IN BOLDFACE.

| | Distortion | PSNR | SSIM | MS-SSIM | IW-SSIM | IFC | VIF | FSIM | GMSD | NLOG-MSE | NLOG-COR | mQGL | sQGL |
|---|---|---|---|---|---|---|---|---|---|---|---|---|---|
| LIVE | JP2K | 0.8954 | 0.9614 | 0.9654 | 0.9653 | 0.9100 | 0.9683 | **0.9717** | **0.9711** | 0.9499 | 0.9515 | 0.9668 | **0.9704** |
| | JPEG | 0.8809 | 0.9764 | 0.9793 | **0.9809** | 0.9440 | **0.9842** | **0.9834** | 0.9782 | 0.9610 | 0.9629 | 0.9800 | 0.9805 |
| | WN | **0.9854** | 0.9694 | 0.9731 | 0.9671 | 0.9377 | 0.9845 | 0.9652 | 0.9737 | **0.9877** | **0.9880** | 0.9591 | 0.9694 |
| | GB | 0.7823 | 0.9517 | 0.9584 | **0.9722** | 0.9649 | **0.9722** | **0.9708** | 0.9567 | 0.9440 | 0.9470 | 0.9543 | 0.9651 |
| | FF | 0.8907 | **0.9556** | 0.9321 | 0.9443 | **0.9644** | **0.9652** | 0.9499 | 0.9416 | 0.9127 | 0.9148 | 0.9482 | 0.9485 |
| CSIQ | AWN | 0.9363 | 0.8974 | 0.9471 | 0.9377 | 0.8460 | 0.9571 | 0.9262 | **0.9676** | **0.9663** | **0.9664** | 0.9561 | 0.9661 |
| | JPEG | 0.8882 | 0.9546 | 0.9622 | **0.9664** | 0.9395 | **0.9705** | 0.9654 | 0.9651 | 0.9483 | 0.9475 | 0.9593 | **0.9661** |
| | JP2K | 0.9363 | 09606 | **0.9691** | 0.9681 | 0.9262 | 0.9672 | 0.9685 | **0.9717** | 0.9503 | 0.9481 | 0.9588 | **0.9722** |
| | PGN | 0.9338 | 0.8922 | 0.9330 | 0.9057 | 0.8279 | **0.9509** | 0.9234 | 0.9502 | **0.9588** | **0.9594** | 0.9393 | 0.9468 |
| | GB | 0.9289 | 0.9609 | 0.9720 | **0.9781** | 0.9593 | **0.9747** | **0.9729** | 0.9712 | 0.9519 | 0.9519 | 0.9546 | 0.9721 |
| | Contrast | 0.8622 | 0.7922 | **0.9521** | **0.9540** | 0.5416 | 0.9361 | 0.9420 | 0.9040 | 0.9238 | 0.9264 | 0.9359 | **0.9434** |
| TID2013 | AWGN | **0.9291** | 0.8671 | 0.8645 | 0.8438 | 0.6611 | 0.8994 | 0.8973 | **0.9462** | 0.9251 | 0.9245 | 0.9109 | **0.9415** |
| | ANMC | **0.8984** | 0.7726 | 0.7729 | 0.7514 | 0.5351 | 0.8299 | 0.8207 | **0.8684** | 0.8414 | 0.8414 | 0.8291 | **0.8641** |
| | SCN | 0.9198 | 0.8515 | 0.8543 | 0.8166 | 0.6601 | 0.8834 | 0.8749 | **0.9350** | 0.9242 | **0.9250** | 0.8958 | **0.9290** |
| | MN | 0.5416 | 0.7767 | 0.8014 | 0.8063 | 0.6732 | **0.8642** | 0.8013 | 0.7075 | **0.8271** | **0.8298** | 0.8027 | 0.7670 |
| | HFN | **0.9141** | 0.8634 | 0.8603 | 0.8553 | 0.7405 | 0.8972 | 0.8983 | **0.9162** | 0.9001 | 0.8993 | 0.8940 | **0.9123** |
| | IMN | **0.8968** | 0.7503 | 0.7628 | 0.7281 | 0.6407 | 0.8536 | 0.8072 | 0.7637 | **0.8799** | **0.8763** | 0.8008 | 0.7392 |
| | QN | 0.8808 | 0.8657 | 0.8705 | 0.8467 | 0.6282 | 0.7853 | 0.8719 | **0.9049** | **0.8917** | 0.8912 | 0.8660 | **0.9018** |
| | GB | 0.9149 | 0.9667 | 0.9672 | **0.9701** | 0.8906 | 0.9649 | 0.9550 | 0.9113 | **0.9694** | **0.9705** | 0.9665 | 0.9543 |
| | DEN | 0.9480 | 0.9254 | 0.9267 | 0.9152 | 0.7779 | 0.8910 | 0.9301 | **0.9525** | **0.9488** | 0.9478 | 0.9352 | **0.9483** |
| | JPEG | 0.9189 | 0.9200 | 0.9265 | 0.9186 | 0.8356 | 0.9191 | 0.9324 | **0.9507** | **0.9553** | 0.9469 | 0.9366 | **0.9475** |
| | JP2K | 0.8840 | 0.9468 | 0.9504 | 0.9506 | 0.9077 | 0.9516 | 0.9576 | **0.9657** | **0.9614** | 0.9598 | 0.9603 | **0.9653** |
| | JGTE | 0.7685 | **0.8493** | 0.8475 | 0.8387 | 0.7425 | 0.8409 | 0.8463 | 0.8403 | 0.8117 | 0.8143 | **0.8605** | 0.8542 |
| | J2TE | 0.8883 | 0.8828 | 0.8888 | 0.8656 | 0.7769 | 0.8760 | 0.8912 | 0.9136 | **0.9371** | **0.9344** | 0.9097 | 0.9176 |
| | NEPN | 0.6860 | 0.7821 | 0.7968 | 0.8010 | 0.5736 | 0.7719 | 0.7917 | **0.8140** | 0.7509 | 0.7554 | **0.8039** | 0.8163 |
| | Block | 0.1552 | 0.5720 | 0.4800 | 0.3716 | 0.2413 | 0.5306 | 0.5489 | **0.6625** | 0.5926 | 0.6148 | **0.6379** | 0.6581 |
| | Mean shift | 0.7672 | 0.7752 | **0.7906** | 0.7833 | 0.5522 | 0.6275 | 0.7530 | 0.7351 | **0.7993** | **0.8009** | 0.7181 | 0.7144 |
| | Contrast | 0.4403 | 0.3775 | 0.4633 | 0.4592 | -0.180 | **0.8385** | 0.4686 | 0.3235 | 0.4654 | 0.4677 | **0.4856** | 0.3423 |
| | CCS | **0.0944** | -0.414 | -0.410 | -0.420 | -0.403 | -0.310 | **-0.275** | **-0.295** | -0.317 | -0.342 | -0.383 | -0.316 |
| | MGN | **0.8905** | 0.7803 | 0.7785 | 0.7727 | 0.6142 | 0.8468 | 0.8469 | **0.8886** | 0.8678 | 0.8676 | 0.8357 | **0.8713** |
| | CN | 0.8411 | 0.8566 | 0.8527 | 0.8761 | 0.8160 | 0.8946 | 0.9120 | **0.9298** | **0.9277** | 0.9245 | 0.8983 | 0.9210 |
| | LCN | 0.9145 | 0.9057 | 0.9067 | 0.9037 | 0.8180 | 0.9203 | 0.9466 | **0.9629** | 0.9339 | 0.9310 | 0.9405 | **0.9625** |
| | CQD | **0.9269** | 0.8542 | 0.8554 | 0.8401 | 0.6006 | 0.8414 | 0.8759 | 0.9102 | **0.9176** | 0.9062 | 0.8924 | **0.9104** |
| | Chr. abr. | **0.8873** | 0.8775 | 0.8784 | 0.8681 | 0.8209 | 0.8848 | 0.8714 | 0.8530 | **0.8872** | **0.8902** | 0.8837 | 0.8619 |
| | Sampling | 0.9042 | 0.9461 | 0.9482 | 0.9474 | 0.8884 | 0.9352 | 0.9565 | **0.9683** | 0.9579 | 0.9573 | 0.9575 | **0.9653** |
| | hit number | 9 | 2 | 3 | 6 | 1 | 8 | 7 | **18** | **16** | 11 | 4 | **20** |

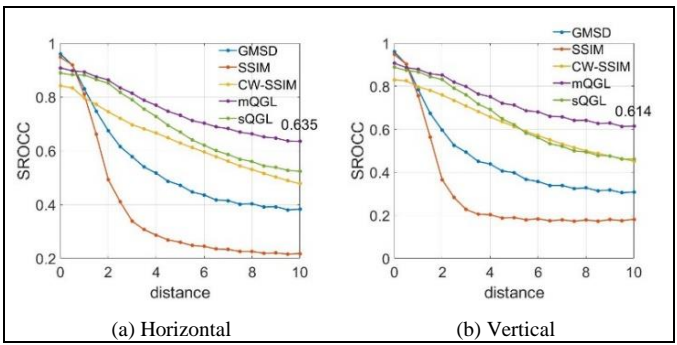

Fig. 5. Comparisons of the SROCC values for several metrics on LIVE database along with the spatial translation distance by pixel. The curves from top to bottom in the legend are: the GMSD method, the SSIM method, the CW-SSIM method, the proposed mQGL with scale factor 1, and the proposed sQGL with scale factor 1 for Gaussian filter.

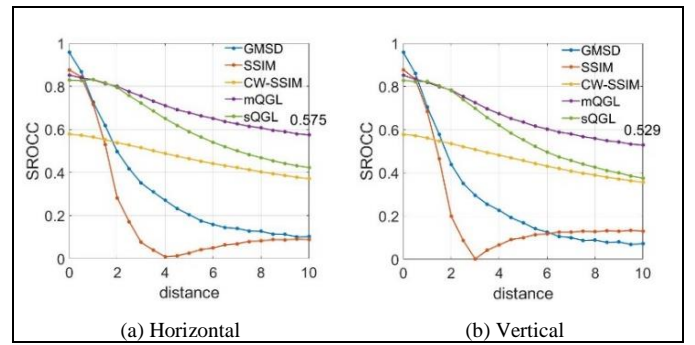

Fig. 6. Comparison of the SROCC values for several metrics on CSIQ database along with the spatial translation distance by pixel. The curves from top to bottom in the legend are: the GMSD method, the SSIM method, the CW-SSIM method, the proposed mQGL with scale factor 1, and the proposed sQGL with scale factor 1 for Gaussian filter.

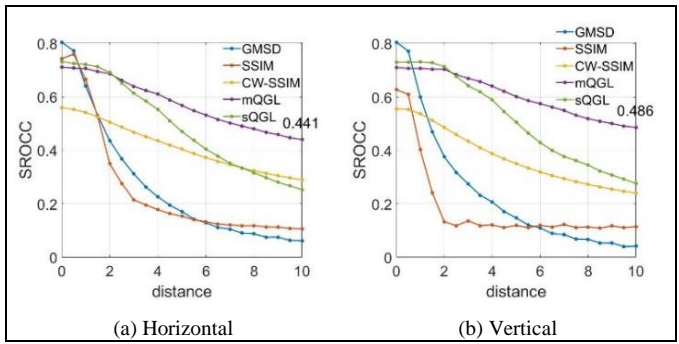

Fig. 7. Comparison of the SROCC values for several metrics on TID2013 database along with the spatial translation distance by pixel. The curves from top to bottom in the legend are: the GMSD method, the SSIM method, the CW-SSIM method, the proposed mQGL with scale factor 1, and the proposed sQGL with scale factor 1 for Gaussian filter.

On CSIQ database, the SROCC of the mQGL metric holds 0.575 and 0.529 when the reference image is translated by 10 pixels along horizontal and vertical directions, while the GMSD and SSIM models drop below 0.2, and CW-SSIM falls below 0.4. Note that the average pooling method shows better performance than the standard deviation based pooling in our computational model, while sQGL is still superior to CW-SSIM in translation situations.

The comparison on TID2013 database also gives the result that the proposed model works better on shift-insensitive property. The SROCC of the proposed mGQL method holds 0.441 and 0.486 when the reference image is translated by 10 pixels along horizontal and vertical directions, while the GMSD and SSIM metrics drop below 0.2, and CW-SSIM falls below 0.3. The average pooling method still shows better performance than the standard deviation pooling method in the shift-insensitive experiment, while sQGL still surpasses the rest models.

These comparisons validates that the proposed average based model mQGL works stably on shift-insensitive property, and is significantly better than the competitors on the three databases when small translation is given to the reference image. This result shows more practical significance since images obtained in many applications suffer dithering or movement.

## IV. DISCUSSION AND CONCLUSION

In this paper, we propose an FR-IQA model based on the local structure similarity of the root of the quadratic sum of the normalized GM and LOG signals. This study proves that the proposed QGL feature is efficient for image quality predication and has the shift-insensitive property. According to this property, the QGL similarity map can measure local quality of the image with a tolerance of image displacement. Furthermore, we compare pooling strategies for the proposed model. The overall image quality computed by standard deviation pooling is highly relevant to subjective image quality compared to the average pooling strategy. Compared with the state-of-the-art FR methods, our proposed model performs excellently and more stably across the three benchmark databases, especially on the largest TID2013 database. The proposed metric with standard deviation pooling also shows stable result on different types of distortions separately on these databases. In summary, the proposed combination strategy of GM and LOG signals is efficient in IQA model design.

Another contribution in this work is that the proposed model shows good performance and robust property on IQA with small spatial translations compared with existing shift-insensitive IQA metric, i.e., CW-SSIM. Experimental results validate that the propose mQGL is superior to CW-SSIM in terms of both the evaluation accuracy and shift-insensitive property. Meanwhile, the proposed mQGL is very simple which only consists of the QGL feature comparison between the reference and distortion images in a single scale, while the CW-SSIM operates on the Gabor filters with a 6-level complex steerable pyramid decomposition on 16 orientations in default, thus the large scale of filtration leads to insensitivity to distortion levels, and the multidirectional calculation increases the computational cost. Obviously, the proposed mQGL would be easier for practical usages compared to CW-SSIM. Nowadays, image error measures for two images still work in premise of alignment of the two images to make sure the point-wise error measure can be carried out exactly. However, practical images obtained from different ways suffer different dithering and movement so that the distorted image is not always totally registered with the reference image, the shift-insensitive property might be more efficient and significant in most practical applications. Furthermore, shift-insensitive FR-IQA metric would be significant in discriminating different types of local textures, for which we will investigate in the future. As for the design of CNNs, the shift-invariance property is desired but is still a pending issue at present, as summarized in the section I. In future, we will further study the feasibility of recognizing translated patterns using QGL features, especially in CNN applications.

In conclusion, with a simple combination of symmetric GM and LOG atoms, the new feature we suggested is efficient in representing image edge features with ability to measure the distortion in images, while insensitive to small spatial translations. Consequently, the proposed model using the QGL feature works robustly in image quality evaluation, and shows obvious advantages by comparison with other outstanding FR-IQA methods when small spatial shift occurs in images. Based on the results, the shift-insensitive QGL feature leads to great significance for practical applications, especially in local texture recognizing and the loss function design in deep neural networks, where the translation-invariance property is considered to be helpful in anti-interference capability.

REFERENCES

[1] Z. Wang, A. C. Bovik and H. R. Sheikh, and E. P. Simoncelli, "Image quality assessment: from error visibility to structural similarity," *IEEE Trans. Image Process.*, vol. 13, no. 4, pp. 600-612, Apr. 2004.
[2] Z. Wang, E. P. Simoncelli, and A. C. Bovik, "Multiscale structural similarity for image quality assessment," in *Proc. IEEE 37th Conf. Rec. Asilomar Conf. Signals, Syst. Comput.*, vol. 2, Nov. 2003, pp. 1398-1402.
[3] Z. Wang, and Q. Li, "Information content weighting for perceptual image quality assessment," *IEEE Trans. Image Process.*, vol. 20, no. 5, pp. 1185-1198, May 2011.

[4] H. R. Sheikh, A. C. Bovik and G. de Veciana, "An information fidelity criterion for image quality assessment using natural scene statistics," *IEEE Trans. Image Process.*, vol.14, no.12, pp. 2117- 2128, Dec. 2005.

[5] H. R. Sheikh. and A.C. Bovik, "Image information and visual quality," *IEEE Trans. Image Process.*, vol.15, no.2, pp. 430- 444, Feb. 2006.

[6] Lin Zhang, Lei Zhang, X. Mou and D. Zhang, "FSIM: A Feature Similarity Index for Image Quality Assessment," *IEEE Trans. Image Process.*, vol. 20, no. 8, pp. 2378-2386, Aug. 2011.

[7] W. Xue, L. Zhang, X. Mou, et al. "Gradient Magnitude Similarity Deviation: A Highly Efficient Perceptual Image Quality Index," *IEEE Trans. Image Process.*, vol. 23, no. 2, pp. 684-695, Feb. 2014.

[8] D. O. Kim, H. S. Han, and R. H. Park, "Gradient information-based image quality metric," *IEEE Transactions on Consumer Electronics*, vol. 56, no. 2, pp. 930-936, 2010.

[9] G. H. Chen, C. L. Yang and S. L. Xie, "Gradient-based structural similarity for image quality assessment," 13$^{th}$ *IEEE International Conference on Image Processing*, 2006.

[10] A. Liu, W. Lin, and Manish Narwaria, "Image Quality Assessment Based on Gradient Similarity," *IEEE Transactions on Image Processing*, vol. 21, no. 4, pp. 1500-1512, 2012.

[11] G. Q. Cheng, J. C. Huang, C. Zhu, Z. Liu, and L. Z. Cheng, "Perceptual image quality assessment using a geometric structural distortion model," in *Proc. 17th IEEE ICIP*, Sep. 2010, pp. 325–328.

[12] L. J. Croner and E. Kaplan. "Receptive Fields of P and M Ganglion Cells Across the Primate Retina." *Vision Research*, Vol. 35, No. 1, pp. 7-24, 1995.

[13] M. Zhang, W. Xue and X. Mou. "Reduced Reference Image Quality Assessment Based on Statistics of Edge." In *Proc. IS&T/SPIE Electronic Imaging*, Vol. 7876, California, USA, 2011.

[14] M. Zhang, X. Mou, and L. Zhang. "Non-shift edge based ratio (NSER): an image quality assessment metric based on early vision features." *IEEE Signal Processing Letters*, vol. 18, no. 5, pp. 315-318, 2011.

[15] W. Shao and X. Mou. "Edge patterns extracted from natural images and their statistics for reduced-reference image quality assessment." In *Proc. IS&T/SPIE Electronic Imaging*, vol. 8660, California, USA, 2013.

[16] W. Xue and X. Mou. "Image quality assessment with mean squared error in a log based perceptual response domain." Signal and Information Processing (ChinaSIP), 2014 *IEEE China Summit & International Conference on IEEE*, pp. 315-319, 2014.

[17] W. Xue, X. Mou, L. Zhang, and A. C. Bovik. "Blind image quality assessment using joint statistics of gradient magnitude and Laplacian features." *IEEE Transactions on Image Processing*, Vol. 23, No. 11, pp. 4850-4862, 2014.

[18] X. Mou, W. Xue, C. Chen, and L. Zhang. "LoG acts as a good feature in the task of image quality assessment." In *Proc. IS&T/SPIE Electronic Imaging*, vol. 9023, California, USA, 2014.

[19] X. Mou, W. Xue, and L. Zhang. "Reduced reference image quality assessment via sub-image similarity based redundancy measurement." In *Proc. IS&T/SPIE Electronic Imaging*, vol. 8291, California, USA, 2012.

[20] Y. Chen, W. Xue, and X. Mou. "Reduced-reference image quality assessment based on statistics of edge patterns." In *Proc. IS&T/SPIE Electronic Imaging*, vol. 8299, California, USA, 2012.

[21] C. Chen, and X. Mou, "A Reduced-Reference Image Quality Assessment Model Based on Joint-Distribution of Neighboring LOG Signals," In *Proc. IS&T Electronic Imaging*, 2016(18): 1-8.

[22] E. P. Simoncelli, and B. A. Olshausen, "Natural image statistics and neural representation." *Annu. Rev. Neurosci*, 24:1193-216, 2001.

[23] B. Zitova, and J. Flusser, "Image registration methods: a survey." *Image & Vision Computing*, vol. 21, no. 11, pp. 977-1000, 2003.

[24] A. Krizhevsky, I. Sutskever, and G. E. Hinton, "ImageNet classification with deep convolutional neural networks." *Advances in Neural Information Processing Systems*, vol. 25, no. 2, 2012.

[25] A. Krizhevsky, I. Sutskever, and G. E. Hinton, "ImageNet classification with deep convolutional neural networks." *Communications of the ACM*, 2017.

[26] M. D. Zeiler and R. Fergus, "Visualizing and understanding convolutional networks." In *European Conference on Computer Vision*. Springer, Cham, 2014

[27] L. A. Gatys, A. S. Ecker, and M. Bethge, "Image style transfer using convolutional neural networks," In *IEEE Computer Vision & Pattern Recognition*, Dec. 2016.

[28] H. Li, Z. Lin, *et al*., "A convolutional neural network cascade for face detection," In *IEEE Computer Vision & Pattern Recognition*, Oct. 2015.

[29] G. Eilertsen, J. Kronander, G. Denes, *et al.*, "HDR image reconstruction from a single exposure using deep CNNs," *ACM Trans. Graph.*, vol. 36, no. 6, Nov. 2017.

[30] G. Wang, Y. Zhang, X. Ye, and X. Mou, "Artificial neural networks," in *Machine Learning for Tomographic Imaging*, IOP Publishing Ltd, Dec. 2019, pp. 3-1 to 3-60. [Online]. Available: https://iopscience.iop.org/book/978-0-7503-2216-4.

[31] H. Zhao, O. Gallo, *et al.*, "Loss functions for image restoration with neural networks," *IEEE Transactions on Computational Imaging*, vol. 3, no.1, pp. 47-57, 2017.

[32] Y. Huang, M. Wang, Y. Qian, *et al.*, "Image completion based on Gans with a new loss function," In 3$^{rd}$ *International Conference on Machine Vision and Information Technology*, Feb. 2019.

[33] H. Huang, H. Tao, and H. Wang, "A convolutional neural network based method for low-illumination image enhancement," In *Proc. of the 2$^{nd}$ Int. Conf. on Artificial Intelligence and Pattern Recognition*, Aug. 2019.

[34] G. Tolias, R. Sicre, and H. Jegou, "Particular object retrieval with integral max-pooling of CNN activations," *Computer Science*, 2015.

[35] A. Giusti, D. C. Cirean, J. Masci, *et al.*, "Fast image scanning with deep max-pooling convolutional neural networks," 2013 *IEEE International Conference on Image Processing*, Melbourne, VIC, 2013, pp. 4034–4038.

[36] M. Cho, J. Sun, O. Duchenne and J. Ponce, "Finding matches in a haystack: a max-pooling strategy for graph matching in the presence of outliers," 2014 *IEEE Conference on Computer Vision and Pattern Recognition, Columbus*, 2014, pp. 2091-2098.

[37] A. Azulay and Y. Weiss, "Why do deep convolutional networks generalize so poorly to small image transformations?" *Journal of Machine Learning Research*, vol. 20, no. 184, pp. 1-25, 2019.

[38] M.P. Sampat, Z. Wang, S. Gupta, A.C. Bovik and M.K. Markey, "Complex wavelet structural similarity: a new image similarity index," *IEEE Transactions on Image Processing*, vol. 18, no. 11, pp. 2385-2401, October, 2009.

[39] H. R. Sheikh, Z. Wang, L. Cormack, and A. C. Bovik. (2005) "Live Image Quality Assessment Database Release 2," [Online]. http://live.ece.utexas.edu/research/quality.

[40] E. C. Larson and D. M. Chandler, "Most apparent distortion: full-reference image quality assessment and the role of strategy," *Journal of Electronic Imaging*, vol. 19, no. 1, March 2010.

[41] N. Ponomarenko, O. Ieremeiev, V. Lukin, K. Egiazarian, L. Jin, J. Astola, B. Vozel, K. Chehdi, M. Carli, F. Battisti, C.-C. Jay Kuo, "Color Image Database TID2013: Peculiarities and Preliminary Results," In *Proceedings of 4th Europian Workshop on Visual Information Processing*, pp. 106-111, Paris, France, 2013.

[42] Lin Zhang, Lei Zhang, and Xuanqin Mou, "RFSIM: a feature based image quality assessment metric using Riesz transforms", in: *Proc. IEEE International Conference on Image Processing*, 2010, Hong Kong.

[43] C. F. Li and A. C. Bovik, "Content-partitioned structural similarity index for image quality assessment," *Signal Process., Image Commun.*, vol. 25, no. 7, pp. 517–526, Aug. 2010.

[44] Z. Wang and X. Shang, "Spatial pooling strategies for perceptual image quality assessment," *IEEE Int. Conf. Image Process.*, Sep. 2006, pp. 2945–2948.

[45] A.K. Moorthy and A.C. Bovik, "Visual importance pooling for image quality assessment," *IEEE J. Special Topics Signal Process*, vol. 3, pp. 193-201, April 2009.

[46] Y. Tong, Hubert Konik, F. A. Cheikh and Alain Tremeau, "Full reference image quality assessment based on saliency map analysis," *Journal of Imaging Science*, vol. 54, no. 3, pp. 30503-30503, 2010.

[47] J. Park, K. Seshadrinathan, S. Lee and A.C. Bovik, "VQpooling: Video quality pooling adaptive to perceptual distortion severity," *IEEE Transactions on Image Processing*, vol. 22, no. 2, pp. 610-620, February 2013.

[48] Ninassi, Alexandre, et al.,"Does where you gaze on an image affect your perception of quality? Applying visual attention to image quality metric," *IEEE International Conference on Image Processing*, vol. 2, 2007.